# Preparation and Characterization of Homogeneous YBCO Single Crystals with Doping Level near the SC-AFM Boundary

Ruixing Liang [a, *], D. A. Bonn [a], W. N. Hardy [a], Janice C. Wynn [b],
K. A. Moler [b], L. Lu [c], S. Larochelle [c], L. Zhou [c], M. Greven [c],
L. Lurio [d], S. G. J. Mochrie [e]

[a] *Dept. of Physics and Astronomy, University of British Columbia, Vancouver, B.C. Canada V6T 1Z1*
[b] *Depts. of Applied Physics and Physics, Stanford University, Stanford, CA 94305, USA*
[c] *Depts. of Applied Physics, Physics and Stanford Synchrotron Radiation Laboratory, Stanford University Stanford CA 94305, USA*
[d] *Dept. of Physics, Northern Illinois University, DeKalb, IL 6011, USA*
[e] *Depts. of Physics and Applied Physics, Yale University, New Haven, CT 06520, USA*

**Abstract**

High-purity and homogeneous $YBa_2Cu_3O_y$ single crystals with carrier doping level near the AFM-SC boundary have been obtained in the oxygen content range between $y = 6.340$ and $6.370$. The crystals are ortho-II phase at room temperature and undergo the orthorhombic to tetragonal transition at about 140°C. They show sharp superconducting transitions, with $T_c$ between 4 and 20 K. $T_c$ changes by 0.8 K when the oxygen content $y$ is changed by 0.001, and is also sensitive to annealing conditions near room temperature, due to the dependence of doping on oxygen ordering correlation lengths. Crystals with oxygen content $y$ lower than 6.345 are non-superconducting.

**1. Introduction**

It is known that the high $T_c$ cuprates are antiferromagnetic insulators when there is no doping of charge carriers in the $CuO_2$ planes. With increasing carrier doping, the antiferromagnetic (AFM) order disappears and gives way to high $T_c$ superconductivity (SC). Studies near the AFM – SC boundary are particularly important because many theories make their sharpest and most characteristic predictions in this very low doping region. For instance, quasi-one-dimensional "stripes" may be realized [1-3] and various staggered current phases have been suggested as possible states for the doped antiferromagnet [4-7]. In spin-charge separation scenarios, some theories predict $h/e$ vortices to be stable in this doping range [8-10], and others predict the existence of topological defects called visons that ensure that vortices remain quantized in units of $h/2e$ [11].

However, a detailed insight into this doping region is far from established. One of the obstacles in this regard is the difficulty of preparing highly homogeneous single crystals at very low doping. In particular, because coexistence of AFM ordering and superconductivity, as well as formation of stripes [1-3], are among the theoretically predicted scenarios, sorting out the intrinsic properties becomes problematic if the samples themselves are prone to chemical inhomogeneity that might produce spatially varying carrier doping and superconducting properties.

---

\* Corresponding author. Phone: +1-604-822-1997; Fax: +1-604-822-4750; E-mail: liang@physics.ubc.ca

Among high $T_c$ cuprates, $YBa_2Cu_3O_y$ (YBCO) is one of the most investigated compounds because it has a stoichiometric cation composition and highly perfect crystals can be grown [12,13]. The carrier doping in YBCO depends on the oxygen content $y$, which varies from 6 to 7. At the oxygen content $y = 7$, where the doping is slightly higher than that required for maximum $T_c$, the so-called ortho-I phase has complete Cu-O chains, along the $b$-direction in the basal plane. Reducing the oxygen content from 7 reduces the carrier doping level and generates oxygen vacancies in the Cu-O chain layer. When the oxygen content $y$ is reduced below a critical value thought to lie between 6.25 and 6.4, the superconductivity disappears and gives way to antiferromagnetic ordering. In principle, one can obtain any $T_c$ value between zero and the maximum 93 K by changing the oxygen content $y$. Because of the high mobility of the chain oxygen ion compared to cations such as $Sr^{2+}$, YBCO is, in principle, relatively easily made homogeneous compared to other high $T_c$ cuprates. It has the further advantage that the doping can be manipulated by simple and reversible changes in the annealing conditions following crystal growth.

Despite these advantages there have been serious issues with homogeneity in YBCO crystals with doping level near the AFM – SC boundary. First of all, the crystals are vulnerable to coexistence of multiple phases. Four different phases have been found to exist in the vicinity of the AFM - SC boundary [14, 15]. They are the tetragonal phase, ortho-I, ortho-II, which has alternately full and empty chains, and the ortho-III* phase, which has an ordering of empty-empty-full chains. The tetragonal phase is stable below $y = 6.28$ and the ortho-II phase is dominant between y = 6.32 and 6.60 [14, 15]. However, between y = 6.28 and $y = 6.32$ where the superconducting $T_c$ reaches zero, a variety of instances of coexisting phases has been observed: the tetragonal phase coexisting with the ortho-II phase in samples made by a deoxygenation process, and ortho-II coexisting with ortho-III* in samples made by an oxygenation process [15]. Also, the ortho-I phase exists in orthorhombic samples that have not been annealed sufficiently below 80°C to allow the other phases to develop. Even if single-phase crystals are obtained it is still difficult to achieve sharp superconducting transition, due to the high sensitivity of $T_c$ to changes in the oxygen content in this doping range. Slow oxygen dynamics at these low oxygen contents further confounds the preparation of homogeneous crystals.

## 2. Preparation

In the present study YBCO single crystals with purity of 99.995% and typical sizes of 0.5 ~ 3 mm wide and 0.02 ~ 0.5 mm thick, were grown by a self-flux method using $BaZrO_3$ crucibles [13].

The oxygen content of the crystals was set to around 6.35 by annealing together with YBCO ceramic pellets under pure oxygen flow at 900 to 930°C for 1 week in a tube furnace with temperature stability of 0.5°C, followed by quenching to room temperature under the protection of nitrogen gas flow. To eliminate the oxygen content inhomogeneity that the quenching process produces at the surfaces, the crystals and pellets were sealed together in a small quartz capsule with a dead volume of less than 0.5 $cm^3$ and then annealed at 570°C for 2 weeks followed by quenching into an ice water bath.

To determine the oxygen content $y$ of the crystals, the ceramic pellets were carefully weighed and annealed in flowing oxygen at 350°C until no further weight change could be observed. The oxygen content was then calculated from the weight change by adopting as the reference value the equilibrium oxygen content $y = 6.993$ at 350°C in 1 atm oxygen [17]. This analysis has a precision ± 0.001 limited by precision of the scale, ±0.01 mg. However, the absolute value of the oxygen content may differ from other works by up to 0.06 due to use of different reference oxygen content values. To verify the absolute accuracy of our weight analysis, iodometric titration, which does not rely on reference oxygen content values, was performed on some ceramic pellets and the results were found to agree with the weight analysis to the precision of the titration, ±0.01.

The resulting crystals with oxygen content near 6.35 are orthorhombic and twinned at room temperature. Twin-free crystals were obtained by applying 10 MPa pressure along the $a(b)$-direction, while the samples were held at temperatures between 100 to 150°C.

Annealing at low temperatures was necessary to stabilize the oxygen ordering, and this was carried out by vacuum-sealing the crystals in a glass tube and placing the tube in a temperature-controlled water bath. The typical annealing condition is 3 weeks at 23°C. Annealing at 30 to 120°C was also carried out to study the dependence of $T_c$ on the annealing temperature.

## 3. Characterization and Discussion

To obtain low-doping crystals with sharp superconducting transitions, it was essential to start with 99.995% pure crystals grown in $BaZrO_3$ crucibles [13]. We could not obtain sharp superconducting transitions with 99.9% pure crystals grown using zirconia crucibles [18]. A possible mechanism for inhomogeneity in less pure samples is that common impurities in the chain site such as Al, Co and Fe are trivalent ions and have higher charge than divalent Cu, causing oxygen anions to cluster around them and produce an inhomogeneous oxygen distribution.

Annealing of the crystals in sealed quartz capsules at 570°C for at least one week was also necessary for obtaining homogeneous crystals. Oxygen diffusion at these low oxygen contents is very slow, as indicated by the difference in physical properties found for samples of the same oxygen content made by deoxygenation and oxygenation processes [15, 16]. A higher annealing temperature might make it easier to achieve good homogeneity, but at temperatures above 600°C a reaction of the crystals with quartz glass was observed.

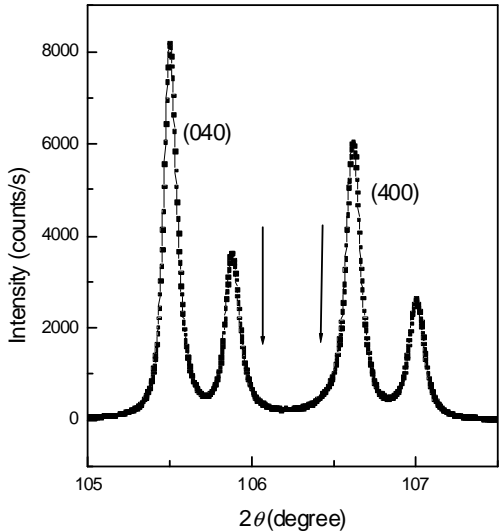

Fig. 1. X-ray $h$-scan for a twinned $y = 6.362$ crystal, measured using the CuK$\alpha$ radiation. The arrows indicate the position of the K$\alpha$1 and K$\alpha$2 lines for the tetragonal phase.

The homogeneous crystals, even with the lowest oxygen content studied in this work, $y = 6.340$, are still orthorhombic at room temperature, although no superconducting transition was found down to 1.5 K. Figure 1 shows the X-ray $h$-scan for a twinned $y = 6.362$ crystal. The splitting between (400) and (040) lines, a result of orthorhombicity, is very clear. The lattice parameters are $a = 0.3843(1)$ nm, $b = 0.3871(1)$ nm and $c = 1.1788(1)$ nm. The orthorhombic to tetragonal transition was found to occur around 140°C, by optical observation of the disappearance of twinning. It should be emphasized that, within the range of oxygen content of the present study, no trace of the tetragonal phase was observed in room temperature X-ray diffraction. Any tetragonal component would have diffraction lines located at the positions indicated by the arrows in Fig. 1. The coexistence of the tetragonal and orthorhombic phases reported in earlier studies [15, 19] must have been caused by inhomogeneity or occurred at oxygen contents lower than the range of the present study, $y = 6.340$.

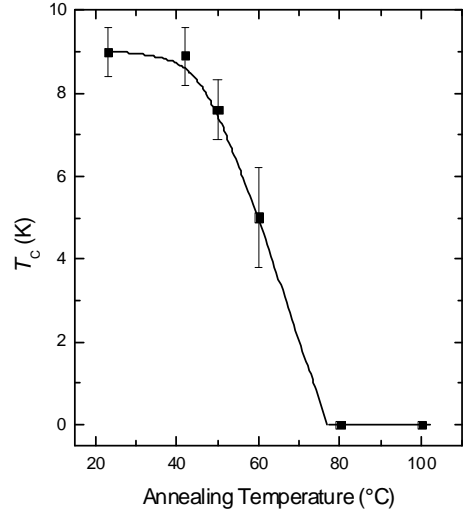

Fig. 2. Saturation $T_c$ for a $y = 6.357$ crystal as a function of the annealing temperature. The crystal became superconducting when annealed at about 60°C. $T_c$ increased rapidly to 9.0 K with decreasing annealing temperature down to 40°C and stays constant with further decrease in annealing temperature.

The occurrence of a slow growth in CuO chain ordering below 80°C is indicated by the time dependence of the superconducting transition temperature $T_c$. The crystals as quenched from 570°C are non-superconducting. After annealing at temperatures below 80°C, superconductivity appears, $T_c$ increases with increasing annealing time and approaches a saturation value, behaviour similar to that reported by Veal et al. [20]. Shown in Fig. 2 is the saturation $T_c$ as a function of annealing temperature. The saturation $T_c$ increases with decreasing annealing temperature, approaching a constant value below about 30°C. The result is reversible and reproducible; that is, the saturation $T_c$ depends only on the final annealing temperature, independent of whether the crystals had been previously annealed at a higher or lower temperature. This indicates that the ordering is thermodynamically stable, and the dependence of $T_c$ on annealing temperature likely results from the temperature dependence of the average length of chain fragments [14,15,20].

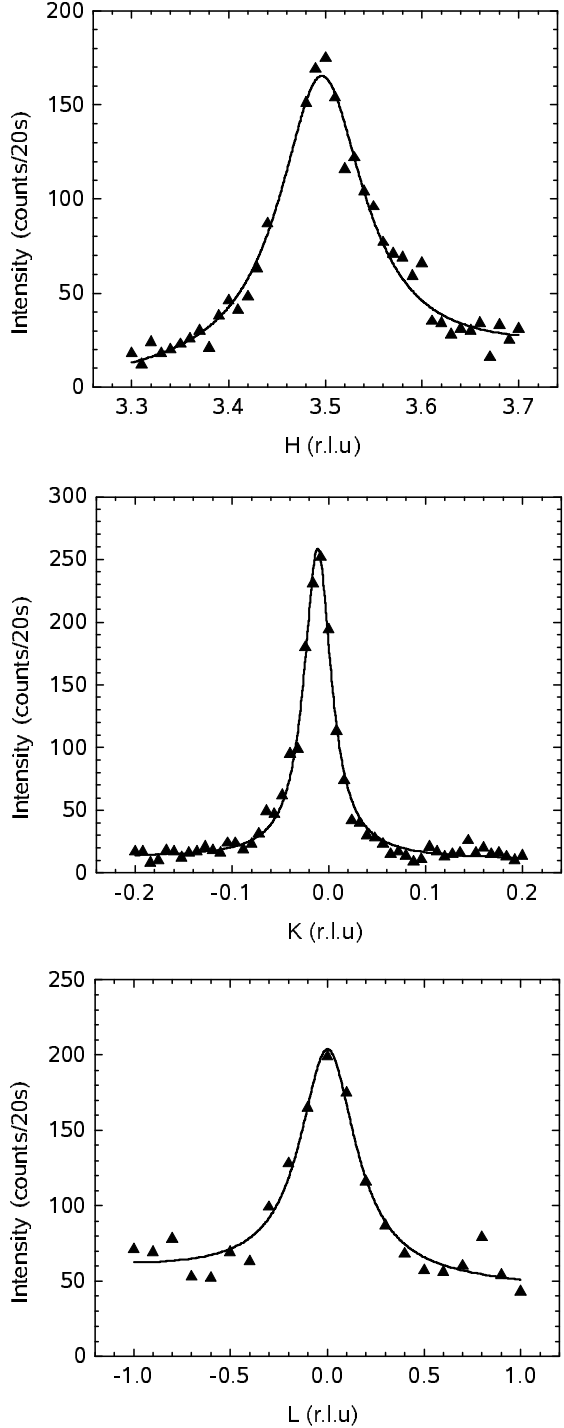

Fig. 3. (3.5, 0, 0) X-ray reflection of ortho-II superlattice measured along *h*, *k* and *l*, for a crystal of *y* = 6.362. Fitting of the profiles to Lorentzian profile yields the correlation lengths $\xi_a$ = 1.14(5) nm, $\xi_b$ = 3.6(3) nm and $\xi_c$ = 1.05(7) nm.

The very low temperature of this ordering suggests that it is associated with an improvement in the degree of ortho-II ordering. Indeed, the present work falls within the reported ortho-II phase region, *y* = 6.30 to 6.60 [14, 15]. Further X-ray diffraction studies, carried out at beamline 8-ID of the Advanced Photon Source, confirmed the existence of (n/2, *k*, *l*) diffraction lines, as shown in Fig. 3, which are unique to the ortho-II phase. Fitting of the line profiles yields oxygen ordering correlation length along *a*, *b*, and *c*-directions of 1.14(5), 3.6(3) and 1.05(7) nm, respectively, for the crystals of *y* = 6.362. No ortho-III* ordering was observed, which was reported to be a metastable phase and exists only in samples made by an oxygenation process [15].

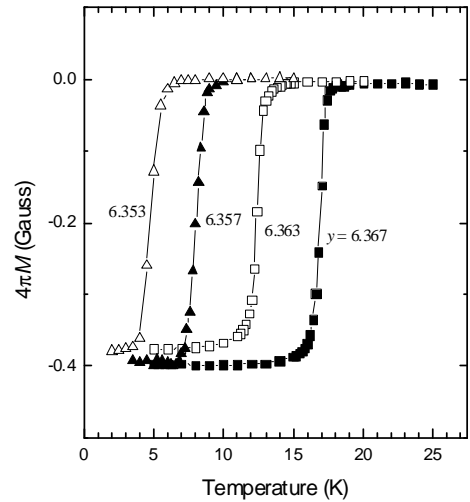

Fig. 4. Field cooling magnetization for crystals after annealing at 23°C. The data were taken in a field of 1 Oe parallel to the *c*-axis. The sharp transition width of 1.0 to 1.6 K indicates a high level of homogeneity.

When cooled in zero field, the superconducting crystals showed full screening of applied magnetic field. In-field cooling typically yields 20% to 60% of complete diamagnetism, depending on the sizes of the crystals, indicating bulk superconductivity. Figure 4 shows the typical in-field cooling magnetization data measure using a SQUID magnetometer at an applied field of 1 Oe parallel to the $c$-axis. The width of the superconducting transition (10% to 90%) is about 1.0 K for the crystals with $T_c > 10$ K, increases slightly with lowering $T_c$, and reaches about 1.6 K for $T_c = 5.5$ K crystals.

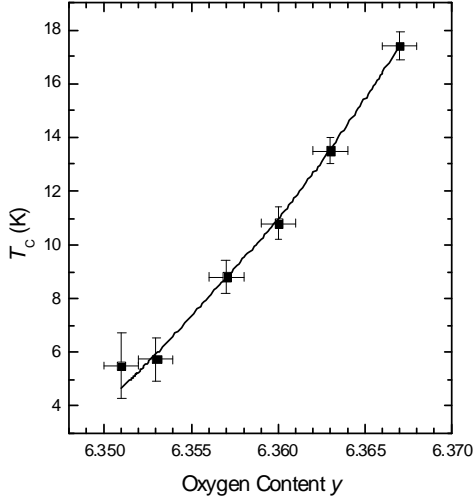

Fig. 5. Superconducting transition temperature $T_c$ as a function of oxygen content, for crystals annealed at 23°C. $T_c$ changes approximately by 0.8 K for a change of 0.001 in oxygen content $y$, which corresponds to about 1°C change in oxygen content setting temperature around 900°C.

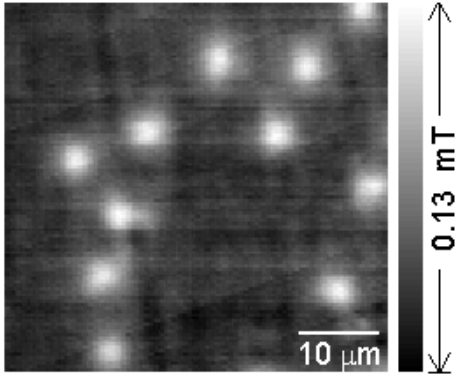

Fig. 6. An image of vortices in a single crystal with $y = 6.361$ ($T_c = 12$ K). The images were taken at 3 K with a scanning Hall probe that had a spatial resolution of better than 2 µm. The degree of uniformity of size and shape of the vortices provides a unique measure of the sample homogeneity, suggesting that the crystal is quite uniform at this length scale.

Figure 5 shows $T_c$ as a function of oxygen content, for crystals annealed at 23°C to saturation values of $T_c$. There is a change in $T_c$ by approximately 0.8 K for a change of 0.001 in oxygen content $y$. This change in $y$ corresponds to a 1°C change in the annealing temperature near 900°C used to initially set the oxygen content. Crystals with oxygen content $y = 6.340$ to 6.345 are non-superconducting, even though they remain orthorhombic at room temperature.

It should be emphasized that, given the steep oxygen content dependence of $T_c$ shown in Fig. 5, the sharp superconducting transition width of 1.0 to 1.6 K shown in Fig. 4 indicates that a high level of oxygen homogeneity was achieved.

Observation of vortices using a scanning Hall probe [21] provided further insight into the homogeneity of the crystals. An image of vortices in a crystal with $y > 6.356$ ($T_c > 8$ K) is shown in Fig. 6. It was found that the vortices are well defined and fairly uniform in size and shape. This provides strong evidence that the superconductivity is bulk in nature. The size of the vortices, which is related to the London penetration depth and thus provides a measure of the doping, was uniform in images taken at different locations in the crystal, indicating that the crystals are uniform on the scale of the probe resolution, about 2 µm.

However, for crystals with y < 6.356 ($T_c < 8$ K) it was found that the vortices are smeared with irregular shapes. This may indicate that for the extremely low doping crystals an even higher degree of homogeneity is needed to form regularly shaped vortices.

## 4. Summary

Homogeneous YBCO crystals with doping levels near the AFM – SC boundary and $T_c$ between 4 and 20 K were obtained in a very narrow oxygen content range between $y = 6.340$ and 6.370. The crystals were found to be single-phase ortho-II and no co-existence of either the tetragonal phase or the ortho-III* was observed.

Low-field magnetization data and images of vortices indicate that the superconductivity of the crystals is bulk in nature. $T_c$ of the crystals is very sensitive to small changes in oxygen content, with a change of 0.001 in oxygen content leading to a change of 0.8 K in $T_c$. Therefore very high homogeneity in oxygen content is essential to obtain sharp superconducting transitions. It was demonstrated that this could be achieved by prolonged annealing of the crystals at 570°C in sealed quartz capsules.

The sensitivity of $T_c$ to low-temperature annealing between 23 and 80°C was studied in some detail. The observed behaviour is consistent with reported temperature dependence of the average Cu-O chain length in the ortho-II phase.

**Acknowledgements**

The Stanford University x-ray group was supported by U.S. Department of Energy under contract Nos. DE-FG03-99ER45773 and DE-AC03-76SF00515, and by NSF CAREER Award No. DMR-9985067. Beamline 8-ID at the Advanced Photon Source is supported by the U.S. Department of Energy via Grant DE-FG02-96ER45593.